\documentclass[useAMS,usenatbib]{mn2e}
\usepackage{psfig}
\newcommand{\co}{${\rm C^{18}O}~J=2\rightarrow 1$}
\newcommand{\rareco}{${\rm C^{17}O}~J=2\rightarrow 1$}
\newcommand{\doublerareco}{${\rm ^{13}C^{17}O}$}

\newcommand{\beq}{\begin{equation}}
\newcommand{\eeq}{\end{equation}}
\begin{document}
\title[Molecular freeze-out in L1689B]
{Molecular gas freeze-out in the pre-stellar core L1689B}
\author[M.P. Redman, J.M.C. Rawlings, D.J. Nutter, D. Ward-Thompson and D.A. 
Williams]
{M.P. Redman$^{1}$, J.M.C. Rawlings$^{1}$, D.J. Nutter$^{2}$,
D. Ward-Thompson$^{2}$, D.A. Williams$^{1}$\\ $^1$ Department of
Physics \& Astronomy, University College London, Gower Street, London
WC1E 6BT, UK.\\ $^2$ Department of Physics and Astronomy, Cardiff
University, PO Box 913, Cardiff CF2 3YB, UK.}
\date{\today}
\pubyear{2001} 
\volume{000}
\pagerange{\pageref{firstpage}--\pageref{lastpage}}
\maketitle 
\label{firstpage}

\begin{abstract}
\rareco\ observations have been carried out towards the 
pre-stellar core L1689B. By comparing the relative strengths of the
hyperfine components of this line, the emission is shown to be
optically thin. This allows accurate CO column densities to be
determined and, for reference, this calculation is described in
detail. The hydrogen column densities that these measurements imply
are substantially smaller than those calculated from SCUBA dust emission 
data. Furthermore, the \rareco\ column densities
are approximately constant across L1689B whereas the SCUBA column
densities are peaked towards the centre. The most likely explanation
is that CO is depleted from the central regions of L1689B. Simple
models of pre-stellar cores with an inner depleted region are compared
with the results. This enables the magnitude of the CO depletion to be
quantified and also allows the spatial extent of the freeze-out to be
firmly established. We estimate that within about 5000 AU of the
centre of L1689B, over 90\% of the CO has frozen onto grains.
This level of depletion can only be achieved after a duration that is at least
comparable to the free-fall timescale.
\end{abstract}


\section{Introduction}
Molecular line profiles from pre-stellar and protostellar objects
potentially offer the best opportunity to extract dynamical
information about the collapse process that leads to the formation of
stars. 
However, it is becoming clear
that the interpretation of these line profiles can be fraught with
difficulty. \citet{rawlings&yates01} used a self-consistent chemical
and dynamical model of collapsing star-forming cores to explore the
effects of abundance variations. They showed that the line profiles
can be very sensitive to the assumed values of the free parameters in
the chemical models. The depletion of molecular species due to
freeze-out can have a profound influence on the line
profiles. Accurate abundances are required deep within these cores in
order that any freeze-out is characterised properly.

The widely used tracer of molecular hydrogen, CO, is so abundant that,
in the cold dark clouds in which stars are forming, it has a large
optical depth and thus cannot trace the densest material. Very
rare CO isotopomers are therefore used and in order of
decreasing abundance these are $^{13}$CO, C$^{18}$O, C$^{17}$O,
$^{13}$C$^{18}$O and \doublerareco, recently discovered by
\citet{bensch.et.al01} toward the $\rho$ Ophiuchi molecular
cloud. Measuring abundances is not straightforward however because
even some of the rare isotopes are mildly optically thick and the
very rare isotopes can require impractically large integration
times. C$^{18}$O is often selected as a result of these conflicting
demands but the analysis can be complicated by the lack of an optical
depth estimate. In contrast, C$^{17}$O is slightly less abundant than
C$^{18}$O and has a complex hyperfine structure revealed
in several distinct line components. For a well
detected line the hyperfine structure can be used to identify whether
optical depth effects are present and hence abundances can
confidently be calculated.

Of course, in order to calculate hydrogen column densities, abundance
ratios between the different isotopomers are
required. \citet{bensch.et.al01} demonstrated how by using several
different isotopes of CO, the optical depths could be cross-checked
and the abundance ratios measured. Effects such as isotope-selective
photo-dissociation and chemical fractionation can be important in
translucent clouds but are not thought to be significant in the cold
dense environments considered here \citep{bensch.et.al01}.

Dust emission measurements can be used to calculate dust column
densities and using a dust to gas ratio, a second estimate for the
column density of hydrogen can be made. If this exceeds the
hydrogen column density as traced by CO then it is possible that the
CO is depleted from the gas-phase. The most probable reason for this
is that the CO has frozen onto the surfaces of
grains. \citet{gibb&little98} find CO abundances reduced by a factor
of at least 10 towards the HH24-26 molecular cloud core by this
method. They also carefully consider alternative explanations that do
not require abundance differences such as optical depth and beam
filling effects and show them to be unlikely. Recently
\citet{bergin.et.al02} have found that not only CO but N$_2$H$^+$ is
depleted in the Bok globule B68. This raises the possibility that in
some cores there will be very few available infall tracers. Two very
recent papers have invesitgated CO depletion. \citet{bacmann.et.al02}
have investigated seven pre-stellar cores and find CO to be
underabundant by factors of 4-15 amongst the
cores. \citet{jorgensen.et.al02} modelled 18 pre-stellar cores and
found that the class 0 sources in their sample are depleted in CO by
an order of magnitude.

L1689B is a pre-stellar core located in
Ophiuchus. \citet{gregersen&evans00} have observed L1689B in the lines
of ${\rm HCO^+}~J=3\rightarrow 2$ and ${\rm H^{13}CO^+}~J=3\rightarrow
2$. Their multipoint data seemed to indicate that the ${\rm
HCO^+}~J=3\rightarrow 2$ emission was double-peaked and blue-skewed, 
indicative of infall. This is supported by the single point ${\rm
CS}~J=2\rightarrow 1$ observations by \citet{lee.et.al99} which show a
clear blue-peak asymmetry. \citet{jessop&wardthompson01} observed
L1689B in C$^{18}$O and although they did not have optical depth
measurements and thus column densities for C$^{18}$O they explored the
dust density and temperature parameter space and argued that the CO has
to be depleted in order to be compatible with the continuum data.

In this paper, observations of the rare ${\rm C^{17}O}$ isotope are
presented and column densities derived and analysed in an attempt to
quantify any depletion that is taking place. These data are discussed
in terms of current models for pre-stellar cores in general and
L1689B in particular.

\section{Observations}
\begin{figure*}
\psfig{file=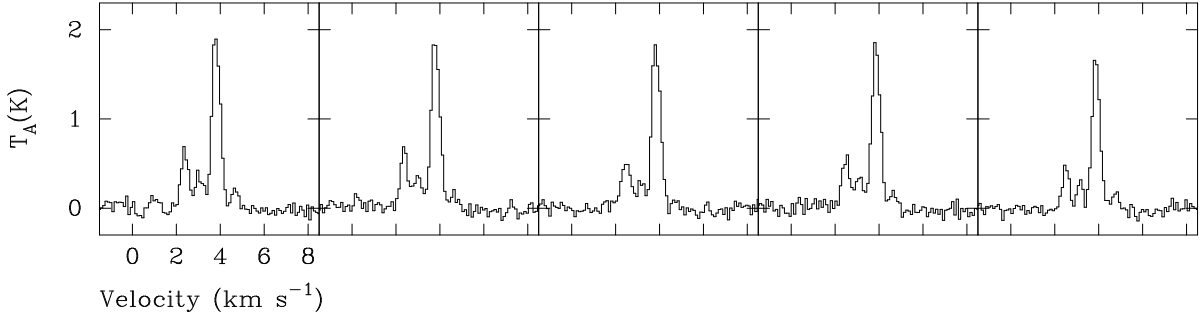,width=450pt,bbllx=43pt,bblly=0pt,bburx=730pt,bbury=183pt}
\caption{\rareco\ line profiles from offset positions of (from left to right) -40\arcsec, -20\arcsec, 0\arcsec, 20\arcsec, 40\arcsec}
\label{fivestrip}
\end{figure*}

The observations were carried out at the James Clark Maxwell Telescope
(JCMT), Mauna Kea, Hawaii on the night of 2001 August 19. The \rareco\
(224.714368 GHz) rotational transition was observed using the the
heterodyne receiver RxA3. The JCMT half-power beam width (HPBW) is 19
arcsec at these frequencies. Typical system temperatures were 325
K. Five pointings at 20 arcsecond offsets at (-40,0), (-20,0), (0,0),
(20,0), (40,0) were obtained for \rareco, centred at the epoch 1950
position (16:31:46.90, -24:31:55.98). \citet{jessop&wardthompson01}
and \citet{gregersen&evans00} used a pointing position of
(16:31:46.98, -24:31:45.00) while \citet{lee.et.al99} used a pointing
position of (16:31:43.6,-24:31:40). The data were reduced in the
standard manner using {\sc specx} and the resulting spectra are
displayed in Figure~\ref{fivestrip}

\section{Results}
\begin{figure}
\psfig{file=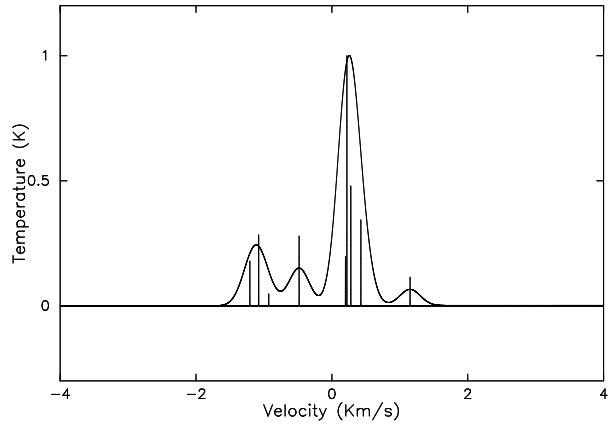,width=225pt,bbllx=11pt,bblly=249pt,bburx=506pt,bbury=599pt}
\caption{Fine structure of the \rareco\ transition. The narrow spikes 
illustrate the relative velocities and strengths of the hyperfine
components. The curve is these components smoothed with a gaussian
profile with $\sigma=0.15~{\rm km~s^{-1}}$. Both the spikes and the
smoothed curve are normalised to their peak intensities}
\label{finestructure}
\end{figure}

\begin{figure}
\psfig{file=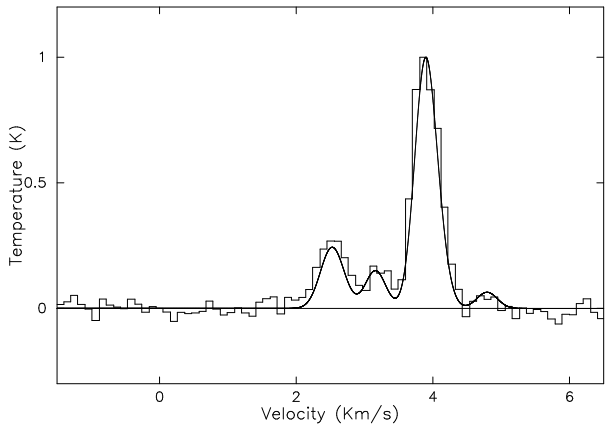,width=225pt,bbllx=15pt,bblly=0pt,bburx=718pt,bbury=498pt}
\caption{\rareco\ data and expected line shape if the emission is optically 
thin with a turbulent velocity dispersion of $0.15~{\rm km~s^{-1}}$ as
in Figure~\ref{finestructure} offset to $3.64~{\rm km~s^{-1}}$. Both
the data and the expected line shape are normalised to their peak
intensities}
\label{datafit}
\end{figure}

Due to the spin of the ${^{17}\rm O}$ nucleus, \rareco\ is composed of
nine hyperfine components. Figure~\ref{finestructure} depicts these
components and their relative strengths (from the tabulation of
\citealt{ladd.et.al98}). The hyperfine components are shown with very 
narrow line widths and also with a broadening of $0.35~{\rm
km~s^{-1}}$ to illustrate the typical line shape that can be expected
to be observed in cold quiescent cores. Since the stronger components
will be affected relatively more by non-negligible optical depths, the
shape of the blended line can be used to verify that the emission is
optically thin in this line and transition. Figure~\ref{datafit}
depicts an overlay of the data from the central position of L1689B
with the expected optically thin line shape. We can conclude from the
close fit that the line does not display any significant saturation
effects and is thus of low optical depth. A detailed fit, using the
HFS routine in the CLASS software package gives optical depths in the
range $0.5-1.5$. For such low optical depths, the actual value
obtained from the software should be treated with some caution as the
deviations from an optically thin line shape are very small (for
example, the fit in Fig.\@~\ref{datafit} is for very low optical
depth) and likely to be strongly affected by the noise. In our later
analysis, we adopt an optical depth of $\sim 0.7$ for all positions
because this is consistent with other recent measurements
(N.J. Evans~{\sc ii}, 2002 private communication). The results of the
fit are given in Table~\ref{taus} which also lists the line centre,
line widths and optical depths obtained.
\begin{table*}
\begin{tabular}{l|l|l|l|l|l|l}
\hline 
Offset & Line centre & Line width & $\tau_{\rm c17o}$ & $\int T_{A}^*~{\rm d}V$ & $N_{\rm c17o}$ & $N_{\rm tot}$\\
       & ${\rm (km~s^{-1})}$ & ${\rm (km~s^{-1})}$ & & ${\rm K~km~s^{-1}}$ & ${\rm cm^{-2}}$ & ${\rm cm^{-2}}$\\
\hline 
$-40\arcsec$ & $3.577\pm 0.004$ & $0.393\pm 0.012$ & $0.673\pm 0.240$ & 1.45 
& $1.0 \times 10^{15}$ & $1.3\times 10^{22}$\\
$-20\arcsec$ & $3.592\pm 0.004$ & $0.378\pm 0.013$ & $1.028\pm 0.242$ & 1.43 
& $9.9 \times 10^{14}$ & $1.3\times 10^{22}$\\
$0\arcsec$ & $3.526\pm 0.004$   & $0.370\pm 0.012$ & $1.346\pm 0.262$ & 1.35 
& $9.3 \times 10^{14}$ & $1.2\times 10^{22}$\\
$20\arcsec$ & $3.608\pm 0.004$  & $0.346\pm 0.013$ & $0.526\pm 0.278$ & 1.34 
& $9.3 \times 10^{14}$ & $1.2\times 10^{22}$\\
$40\arcsec$ & $3.537\pm 0.004$  & $0.356\pm 0.011$ & $1.462\pm 0.273$ & 1.09 
& $7.5 \times 10^{14}$ & $9.9\times 10^{21}$\\
\hline
\end{tabular}
\caption{Optical depths, ${\rm C^{17}O}$ column densities and estimated total column densities for each 
offset position.}
\label{taus}
\end{table*}
The net broadening (due to turbulent and instrumental broadening),
$V_{\rm broad}\simeq 0.37~{\rm km~s^{-1}}$. This is smaller than the
${\rm C^{18}0}$ line width than measured by
\citet{jessop&wardthompson01} who find $V_{\rm FWHM}\simeq
0.6-0.8~{\rm km~s^{-1}}$ for their singly peaked and gaussian line
profiles (though as discussed below, this line is actually optically
thick).

\subsection{CO column densities}
\if{wib} Empirical formulae are often used to derive column densities but it is
worthwhile to briefly review them from first principles. If the system
is in local thermodynamic equilibrium then the Boltzmann equation can
be used to describe the level populations and to define the excitation
temperature
\beq 
\frac{n_u}{n_l}=\frac{g_u}{g_l}\exp\left( -\frac{h\nu}{kT_{\rm ex}}\right),
\label{boltzman}
\eeq
where the $n$ terms are the number density of molecules in the given
state and the $g$ terms are the statistical weights. The equation of
radiative transfer can be written as the detection equation
\beq
T_{\rm A}^*=\eta_{\rm B}\left[ J(T_{\rm ex})-J(T_{\rm cmb})\right] 
\left[ 1 - \exp(-\tau)\right],
\label{ta}
\eeq
where
\beq
J(T)=\frac{(h\nu/k)}{[\exp(h\nu/kT) - 1]}.
\label{jt}
\eeq
$T_{\rm A}^*$ is the antenna temperature, $\eta_{\rm B}$ is the beam
efficiency, $T_{\rm cmb}$ is the background temperature. $\tau$ is the
optical depth of the rotational transition between the upper, $u$, and
lower, $l$, states and is given by
\begin{eqnarray}
\tau &=& \left( n_l B_{lu} - n_u B_{ul} \right) \frac{h\nu}{4\pi} 
\frac{L}{\Delta\nu}\\
&=& \frac{c^2}{8\pi\nu^2\Delta\nu} N_l \frac{g_u}{g_l}A_{ul}
\left[ 1-\exp\left(-\frac{h\nu}{kT_{\rm ex}}\right)\right].
\end{eqnarray}
Here, $B$ and $A$ are the Einstein coefficients of absorption and
emission respectively; $L$ is the path length and so $N=nL$ is the
column density; $\Delta\nu\equiv\nu/c\Delta V$ is the line
width. Using the Boltzmann equation
\beq
\tau = \frac{c^2}{8\pi\nu^2\Delta\nu} N_u A_{ul}
\left[ \exp \left( \frac{h\nu}{kT_{\rm ex}}\right) -1 \right],
\eeq
and using the definition of $A_{ul}$
\beq
A_{ul} = \frac{64\pi^4\nu^3}{3hc^3}\frac{\mu^2 S}{g_u},
\eeq
where $\mu$ is the permanent dipole moment and $S$ is the line strength gives
\beq
\frac{N_u}{g_u}=\frac{3h}{8\pi^3} \frac{1}{\mu^2 S}\frac{\tau\Delta V}
{\left[ \exp\left({h\nu}/{kT_{\rm ex}}\right) -1 \right]}. 
\eeq

The column density in the transition is related to the total column
density of the species by
\beq
\frac{N_u}{g_u}=\frac{N_{\rm tot}}{Q(T_{\rm rot})}
\exp\left(-\frac{E_u}{T_{\rm rot}} \right)
\eeq 
where $Q(T_{\rm rot})$ is the partition function and $T_{\rm rot}$ is
the rotational temperature, assumed to be the same for all levels. It
is usually further assumed that $T_{\rm rot}=T_{\rm ex}$ so that
\beq
N_{\rm tot}=\frac{3h}{8\pi^3}\frac{Q(T_{\rm ex})}{\mu^2 S}\tau
\Delta V\frac{\exp({E_u}/{k T_{\rm ex}})}{\exp({h\nu}/{kT_{\rm ex})-1}}.
\label{ntotv1}
\eeq
This form of the column density can be used if $\tau$ and $\Delta V$
are known but requires a knowledge of $T_{\rm ex}$.

Alternatively, if the optical depth is small then Eqn~(\ref{ta}) becomes
\beq
T_{\rm A}^*/{\eta_{\rm B}}\simeq\left[ J(T_{\rm ex})-J(T_{\rm cmb}) \right]\tau
\eeq
and substituting for $\tau$ in Eqn~(\ref{ntotv1}) and using Eqn~(\ref{jt}) 
gives
\begin{eqnarray}\nonumber
N_{\rm tot}&=&\frac{3k}{8\pi^3\nu}\frac{Q(T_{\rm ex})}{\mu^2 S}
\frac{J(T_{\rm ex})} {J(T_{\rm ex})-J(T_{\rm cmb})}\\
& \times &\exp\left(\frac{E_u}{k T_{\rm ex}}\right)\int T_{\rm mb}{\rm d}V
\end{eqnarray}
where we have also used ${T_{\rm A}^*\Delta V/{\eta_{\rm B}}}
\simeq \int T_{\rm mb}{\rm d}V$, the integrated line intensity.
Finally, if $T_{\rm ex}>>T_{\rm cmb}$ this can be simplified further
to
\beq
N_{\rm tot}=\frac{3k}{8\pi^3\nu}\frac{Q(T_{\rm ex})}{\mu^2 S}
\exp{\left(\frac{E_u}{k T_{\rm ex}}\right)}\int T_{\rm mb}{\rm d}V.
\label{ntotv2}
\eeq
If the integrated line strength is measured in ${\rm km~s^{-1}}$,
frequency in GHz, $\mu$ in 
debye ($1~{\rm debye} = (1/3)\times 10^{-29}{\rm C~m}$) and if an
SI to CGS conversion factor of $4\pi\epsilon_0$ is used then the
column density is 
\begin{eqnarray}\nonumber
N_{\rm tot}&=&1.67\times 10^{14}~{\rm cm^{-2}}~\frac{Q(T_{\rm
ex})}{\nu\mu^2 S}\\&\times &\exp{\left(\frac{E_u}{k T_{\rm
ex}}\right)}\int T_{\rm mb}{\rm d}V.
\end{eqnarray}
\fi

If it is assumed that the optical depth is small, local thermodynamic
equilibrium holds, and the excitation temperature $T_{\rm ex}>>T_{\rm
cmb}$ then the total column density of a species is given by
\beq
N_{\rm tot}=\frac{3k}{8\pi^3\nu}\frac{Q(T_{\rm ex})}{\mu^2 S}
\exp{\left(\frac{E_u}{k T_{\rm ex}}\right)}\int T_{\rm mb}{\rm d}V.
\label{ntotv2}
\eeq
where $\int T_{\rm mb}{\rm d}V\simeq{T_{\rm A}^*\Delta V/{\eta_{\rm
B}}}$, the integrated line intensity. If the integrated line strength
is measured in ${\rm km~s^{-1}}$, frequency in GHz, $\mu$ in debye
($1~{\rm debye} = (1/3)\times 10^{-29}{\rm C~m}$) and if an SI to CGS
conversion factor of $4\pi\epsilon_0$ is used then the column density
is
\begin{eqnarray}\nonumber
N_{\rm tot}&=&1.67\times 10^{14}~{\rm cm^{-2}}~\frac{Q(T_{\rm
ex})}{\nu\mu^2 S}\\&\times &\exp{\left(\frac{E_u}{k T_{\rm
ex}}\right)}\int T_{\rm mb}{\rm d}V.
\end{eqnarray}
The excitation temperature is taken to be 12K, as inferred by
\citet{jessop&wardthompson01}.  The rotational energy levels are
assumed to be thermalized at $T_{\rm ex}$ and the partition function
$Q(T_{\rm ex})$ for each species is evaluated by interpolating between
values obtained from the JPL molecular line database
\citep{pickett.et.al98} and these are also listed in Table
2. Calculating a hydrogen column density from the ${\rm C^{17}O}$
emission requires knowledge of three conversion factors: the ${\rm
H_2}/{\rm CO}$ abundance ratio, which we take to be $3700$
\citep{lacy.et.al94}; the oxygen isotope ratios ${\rm ^{16}O}/{\rm
^{18}O}$, taken to be 560 \citep{wilson&rood94} and ${\rm ^{18}O}/{\rm
^{17}O}$, taken to be 3.65
\citep{ladd.et.al98,penzias81}. The final conversion factor from
${\rm C^{17}O}$ to ${\rm H_2}$ is $7.56\times 10^{6}$.

\begin{table}
\begin{tabular}{c|c|c}
\hline 
Parameter & $\rm C^{17}O~J=2\rightarrow 1$ & $\rm
C^{18}O~J=2\rightarrow 1$ \\
\hline 
$\nu$~(GHz) & 224.714 & 219.560\\
$E_u$~(K)& 16.177& 15.8058\\
$\mu$~(Debye)& 0.11034& 0.11079\\
$S$ & 2& 2\\
$Q(12.0~{\rm K})$ & 4.80 & 4.93\\
\hline
\end{tabular}
\caption{Molecular data for $\rm C^{17}O$ and $\rm C^{18}O$.}
\end{table}

Column densities are also calculated using archive \co\ data of
\citet{jessop&wardthompson01}. This is possible since the \rareco\
line is optically thin and so the optical depth of the \co\ line can
be calculated. \citet{ladd.et.al98} show that simply using the ratio
of the peaks in these lines to calculate the optical depth is
unsuitable. This is because the closely spaced fine structure of the
\rareco\ line means that the peak in this transition is very sensitive
to the turbulent velocity. The integrated intensities are much more
robust and the ratio of these are compared with figure 1 of
\citet{ladd.et.al98} to infer that the optical depth in the \co\ line 
is $\tau_{C18O}\ga 2$. When the optical depth is moderate as in
this case, Eqn~(\ref{ntotv2}) multiplied by a correction factor of
$\tau/(1-\exp{-\tau})$ can be used to calculate column density.

\begin{figure}
\psfig{file=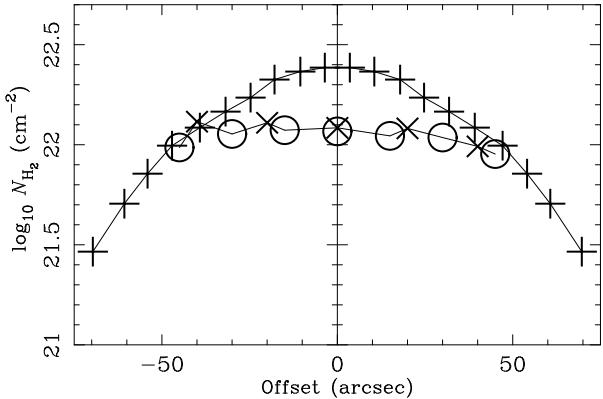,width=225pt,bbllx=38pt,bblly=0pt,bburx=679pt,bbury=442pt}
\caption{Column densities across L1689B, measured using the ${\rm C^{17}O} 
J=2\rightarrow 1$ transition (crosses), ${\rm C^{18}O}$ data of
\citet{jessop&wardthompson01} (circles) and from SCUBA $850~\mu{\rm
m}$ radial profile fits of \citet{evans.et.al01} (pluses). The sizes
of the symbols may be taken as a rough indication of the observational
error.}
\label{columns}
\end{figure}

The column densities derived from the two lines are plotted as a
function of offset in Figure~\ref{columns}. The agreement is excellent
and both datasets show that the column density is remarkably flat
across the face of the core.

\subsection{Dust column densities}
\citet{evans.et.al01} 
obtained SCUBA maps of L1689B and carried out a detailed radiative
transfer analysis of the dust emission which enable them to place
constraints on the temperature and density distribution in the
envelope. The gas density distribution was well fitted by a
Bonner-Ebert sphere with a central density of $n_{\rm c}=1\times
10^6~{\rm cm^{-3}}$ and outer radius of $3\times 10^4~{\rm AU}$.
The dust opacities used were from \citet{ossenkopf&henning94}. Using the
azimuthally averaged results of \citet{evans.et.al01}, and their dust
opacity, we calculate dust column densities, based on an assumed gas
to dust mass ratio of 100 according to the usual formula
\beq
N=S_{850}/[\Omega\kappa_{850} \mu m_{\rm H}
B_{850}(T)]
\eeq
where $S_{850}=0.7~{\rm Jy}$ is the flux at 850 microns within a
$33\arcsec$ beam, $\Omega$ is the solid angle subtended by the beam,
$\kappa_{850\mu{\rm m}}=0.018~{\rm cm^2~g^{-1}}$ is the dust opacity,
$\mu=2.2$ is the mean molecular weight, $m_{\rm H}$ is the mass of
hydrogen and $B_{850\mu{\rm m}}(T)$ is the Planck function at
$T=12~{\rm K}$. The dust column densities are plotted in
Figure~\ref{columns} and are consistent with those obtained by
\citet{bacmann.et.al00,bacmann.et.al02} (who use a different opacity
to calculate their column densities) and \citet{ward-thompson.et.al02}.

\subsection{Depletion of CO}
It is clear from Figure~\ref{columns} that the two independent
techniques for measuring the column density produce results that
differ both in magnitude and distribution. As inferred by the above
discussions, the exact normalisations of the curves in Figure~\ref{columns}
are subject to variation depending on the adopted values of
dust opacity, isotopic ratios and molecular abundances. The most
important aspect of Figure~\ref{columns} is the flatness of the column
density plot from CO compared with the centrally peaked dust
measurements. This strongly indicates that depletion is occurring
because the column density should rise for positions that approach the
dense centre of the object. The most likely reason for this depletion
is that the CO has frozen to the surface of the dust grains deep
within the core. The best fit dust models of \citet{evans.et.al01}
have a dust temperature profile that drops from $\sim 12~{\rm K}$ at
the edge of this pre-stellar core down to a temperature of $\sim
7~{\rm K}$ at the centre. The combination of the higher gas density and
the lower dust temperature is the likely reason for a substantial
freeze-out of CO in the central regions of L1689B. An alternative
explanation for these data is that the CO lies along the line of sight
and is not actually part of the L1689B core. This can be ruled out
since this would require the core to be completely depleted of CO in
order that the column density remains roughly constant across the face
of it.

\begin{figure}
\psfig{file=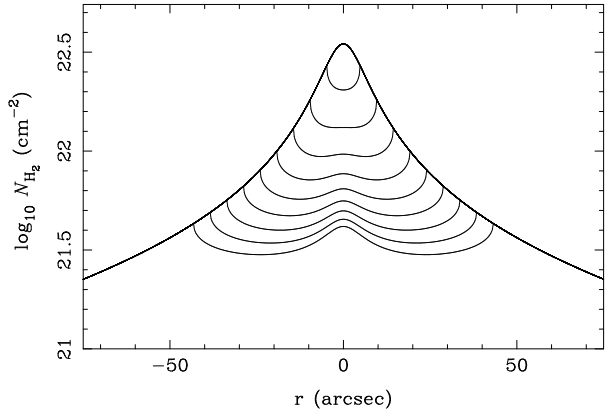,width=225pt,bbllx=16pt,bblly=0pt,bburx=690pt,bbury=463pt}
\caption{
Column densities from a Plummer-like sphere density distribution with
a depletion `hole' that extends from the centre to a radius $R_{\rm
freeze}$. The local depletion within the hole is such that only 5\% of
the CO remains in the gas phase. The upper most curve is with no
depletion. The curves below it are for values of $R_{\rm freeze}$ of
between 5 and 45 arcseconds. The curves have not been convolved with a
telescope beam and hence are more centrally peaked than the
observational data.}
\label{hole}
\end{figure}

\citet{jessop&wardthompson01} 
compared their \co\ data with earlier mm continuum data of
\citet{andre.et.al96}. Using a model for L1689B with power law
profiles for the density, temperature and CO abundance, they explored
the parameter space that would be consistent with both datasets. They
conclude that the CO could be depleted by up to $95\%$ but they did
not have an optical depth measurement for their ${\rm C^{18}O}$ data. In
contrast, our results indicate that the ratio of the column densities
towards the centre is only around a factor of 3 or so.
Of course, the CO
is unlikely to be uniformly depleted and this factor of 3 represents
the column depletion. The local depletion is likely to be much
higher. This can be investigated by plotting the column densities that
result from models where the radial density profile, abundance and
temperature are all varied. To illustrate this in a very simple case,
we use a constant temperature model and vary the abundance such that
interior to a certain radius $R_{\rm freeze}$ the CO is depleted by a
factor 0.95 due to freeze-out. Instead of using the Bonner-Ebert
density distributions which are inconvenient to work with, we use the
parameters of the fit of
\citet{evans.et.al01} in an equivalent Plummer-like sphere which has
a density profile of the form (see,
e.\@g.\@~\citealt{whitworth&wardthompson01,whitworth&bate02})
\begin{equation}
\rho(r)=\frac{\rho_0R_0^2}{(R_0^2+r^2)}
\end{equation}
where $\rho_0=1.0\times 10^6~{\rm cm^{-3}}$ is the central density and
$R_0=750~{\rm AU}$ is an inner radius within which the density is
approximately $\rho_0$. For the purposes here, the difference between
the Bonner-Ebert and Plummer density distributions are of little
consequence - both approximate $\rho\sim r^{-2}$ in the envelope and
are roughly constant close to the centre. Figure~\ref{hole} shows the
column density that results from this simple Plummer sphere model for
a range of values of $R_{\rm freeze}$ with the top curve showing the
underlying density distribution.  Comparison with Figure~\ref{columns}
shows clearly that values of $R_{\rm freeze}$ of about
$40\arcsec\equiv 5000~{\rm AU}$ would be able to reproduce the
approximately flat CO abundance with a column depletion factor of
about the correct degree. One could of course investigate
the parameter space fully and include realistic variations in the
freeze-out and temperature but we defer that to a later paper.

The close fit of the \rareco\ lines by a model with only turbulent
velocities (i.e. with no infall or outflow) indicates that the CO
emitting material is static. In contrast, the ${\rm
HCO^+}~J=3\rightarrow 2$ and ${\rm H^{13}CO^+}~J=3\rightarrow 2$ data
of \citet{gregersen&evans00} show double peaked line profiles
indicating both that the HCO$^+$ is self-absorbed and that the gas is
undergoing bulk gas motions. These emissions are likely to originate
from within the denser regions of the core. The flatness of the
abundance points across the core measured using \rareco\
(Figure~\ref{columns}) indicates that the CO is depleted from the
central regions. Thus one can picture that L1689B is composed of a
dynamical dense core which is depleted in CO but in which HCO$^+$ is
present. This is surrounded by a quiescent surrounding region where
both CO and HCO$^+$ are present.

\section{Conclusions}
Rare isotopes of CO have been observed towards the pre-stellar core
L1689B. By using the hyperfine structure of ${\rm
C^{17}O}~J=2\rightarrow 1$ the transition is confirmed to be optically
thin and emitted by quiescent gas. This allows an estimate of the
total column density as traced by this molecule to be made. A
comparison of this value with that inferred by SCUBA dust emission
measurements reveal the CO to be depleted in the central regions of
this object. The magnitude and extent of the depletion is estimated by
comparison with a simple model of a pre-stellar core with an inner
depleted region. We estimate that within 5000 AU of the centre of
L1689B, around 90\% of the CO has frozen onto grains. The dust
temperature at this radius is $\simeq 10~{\rm K}$ (from figure 5 of
\citealt{evans.et.al01}). The sublimation temperature of CO is 
$\simeq 20~{\rm K}$  so potentially freeze-out could
occur in the outer regions of the cloud. In practice of course, the
timescale for freeze-out in the low density outer regions is long and
freeze-out will occur only in those portions of the cloud with a high
local density and low dust temperature.

For the physical parameters of L1689B, the rate of freeze-out of CO is
given by \citep{rawlings.et.al92}
\beq
\dot{n}_{\rm CO}=4.57\times 10^4 d_{\rm g} a^2 T^{1/2}C n_{\rm H} 
S_{\rm CO}m_{\rm CO}~{\rm cm^{-3}~s^{-1}}
\eeq
where $n_{\rm H}$ and $n_{\rm CO}$ are the hydrogen nucleon and CO
densities respectively, $m_{\rm CO}=28~{\rm amu}$ is the molecular
mass of CO, $d_{\rm g}$ is the ratio of the number density of grains
to CO molecules, $a$ is the grain radius, $C$ is a factor which
accounts for electrostatic effects, $S_{\rm CO}=1$ is the assumed
sticking co-efficient for CO. Using $<d_{\rm g} a^2>\simeq 2.2\times
10^{-22}~{\rm cm^{-2}}$ and $T\sim 10~{\rm K}$ yields
\begin{equation}
\dot{n}_{\rm CO}=6.0\times10^{-18}n_{\rm H}n_{\rm CO}{\rm ~~cm}^{-3}{\rm 
s}^{-1}
\end{equation}
Using a value of $n_{\rm H}=2.8\times10^5$cm$^{-3}$ for L1689B
\citep{bacmann.et.al00} we find that in the absence of other CO
formation and destruction mechanisms, 90\% depletion of CO is achieved
after $\sim$43,400 years.  This is approximately half of the nominal
free-fall timescale (97,400 years) for the value of $n_{\rm H}$ quoted
above, but we should be aware that the freeze-out timescale is a {\it
lower} limit and would be larger if the sticking coefficient were less
than unity, or if (as would seem likely) the cores have condensed from
a less dense state.  We may therefore conclude that the level of CO
depletion may provide a sensitive indicator of the age of cores
relative to their free-fall times and that a C$^{17}$O survey of
sources at various (early) stages of evolution could provide a
powerful diagnostic of their dynamical status.

\section*{Acknowledgements}
We thank the referee for a prompt report that led to an improved
paper. We thank W-F.D. Thi for useful discussions. MPR and DJN are
supported by PPARC. We thank the staff of the JCMT for their excellent
assistance during the observations. The JCMT is operated by the JAC,
Hawaii, on behalf of the UK PPARC, the Netherlands NWO, and the
Canadian NRC. We have made use of the JCMT data archive at the CADC,
which is operated by the Dominion Astrophysical Observatory for the
National Research Council of Canada's Herzberg Institute of
Astrophysics.

\label{lastpage}

\begin{thebibliography}{}


\bibitem[\protect\citeauthoryear{Andr\'{e}, Ward-Thompson \& Motte}{Andr\'{e}
  et~al.}{1996}]{andre.et.al96}
Andr\'{e} P.,  Ward-Thompson D.,    Motte F.,  1996, A\&A, 314, 625

\bibitem[\protect\citeauthoryear{Bacmann, Andr\'{e}, Puget \& et al}{Bacmann
  et~al.}{2000}]{bacmann.et.al00}
Bacmann A.,  Andr\'{e} P.,  Puget J.~L.,    et al 2000, A\&A, 361, 555

\bibitem[\protect\citeauthoryear{Bacmann, Lefloch, Cecarelli \& et al}{Bacmann
  et~al.}{2002}]{bacmann.et.al02}
Bacmann A.,  Lefloch B.,  Cecarelli C.,    et al 2002, A\&A, 389, L6

\bibitem[\protect\citeauthoryear{Bensch, Pak, Wouterloot, Klapper \&
  Winnewisser}{Bensch et~al.}{2001}]{bensch.et.al01}
Bensch F.,  Pak I.,  Wouterloot J. G.~A.,  Klapper G.,    Winnewisser G.,
  2001, ApJ, 562, L185

\bibitem[\protect\citeauthoryear{Bergin, Alves, Huard \& Lada}{Bergin
  et~al.}{2002}]{bergin.et.al02}
Bergin E.~A.,  Alves J.,  Huard T.~L.,    Lada C.~J.,  2002, ApJ, In press


\bibitem[\protect\citeauthoryear{{Evans}, {Rawlings}, {Shirley} \&
  {Mundy}}{{Evans} et~al.}{2001}]{evans.et.al01}
{Evans} N.~J.,  {Rawlings} J.~M.~C.,  {Shirley} Y.~L.,    {Mundy} L.~G.,  2001,
  ApJ, 557, 193

\bibitem[\protect\citeauthoryear{Gibb \& Little}{Gibb \&
  Little}{1998}]{gibb&little98}
Gibb A.~G.,  Little L.~T.,  1998, MNRAS, 295, 299

\bibitem[\protect\citeauthoryear{Gregersen \& Evans}{Gregersen \&
  Evans}{2000}]{gregersen&evans00}
Gregersen E.~M.,  Evans N.~J.,  2000, ApJ, 538, 260

\bibitem[\protect\citeauthoryear{Jessop \& Ward-Thompson}{Jessop \&
  Ward-Thompson}{2001}]{jessop&wardthompson01}
Jessop N.~E.,  Ward-Thompson D.,  2001, MNRAS, 323, 1025

\bibitem[\protect\citeauthoryear{J{\o}rgensen, Sch\"{o}ier \& van Dishoeck}{J{\o}rgensen et~al.}
{2002}]{jorgensen.et.al02}
J{\o}rgensen J.~K., Sch\"{o}ier F.L. van Dishoeck E.~F.,  2002, A\&A, in press

\bibitem[\protect\citeauthoryear{Lacy, Knacke, Geballe \& Tokunaga}{Lacy
  et~al.}{1994}]{lacy.et.al94}
Lacy J.~H.,  Knacke R.,  Geballe T.~R.,    Tokunaga A.~T.,  1994, ApJ, 428, L69

\bibitem[\protect\citeauthoryear{Ladd, Fuller \& Deane}{Ladd
  et~al.}{1998}]{ladd.et.al98}
Ladd E.~F.,  Fuller G.~A.,    Deane J.~R.,  1998, ApJ, 495, 871

\bibitem[\protect\citeauthoryear{Lee, Myers \& Tafalla}{Lee
  et~al.}{1999}]{lee.et.al99}
Lee C.~W.,  Myers P.~C.,    Tafalla M.,  1999, ApJ, 526, 788

\bibitem[\protect\citeauthoryear{Ossenkopf \& Henning}{Ossenkopf \&
  Henning}{1994}]{ossenkopf&henning94}
Ossenkopf V.,  Henning T.,  1994, A\&A, 291, 943

\bibitem[\protect\citeauthoryear{Penzias}{Penzias}{1981}]{penzias81}
Penzias A.~A.,  1981, ApJ, 249, 518

\bibitem[\protect\citeauthoryear{Pickett, Poynter, Cohen, Delitsky \&
  Muller}{Pickett et~al.}{1998}]{pickett.et.al98}
Pickett H.~M.,  Poynter R.~L.,  Cohen E.~A.,  Delitsky M.~L.,    Muller J. C.
  P. H. S.~P.,  1998, J. Quant. Spectrosc. \& Rad. Transfer, 60, 883

\bibitem[\protect\citeauthoryear{Rawlings, Hartquist, Menten \& 
  Williams}{Rawlings et~al.}{1992}]{rawlings.et.al92}
Rawlings J. M.~C., Hartquist T.~W., Menten K.~M., Williams D.~A., 
1992, MNRAS, 255, 471

\bibitem[\protect\citeauthoryear{Rawlings \& Yates}{Rawlings \&
  Yates}{2001}]{rawlings&yates01}
Rawlings J. M.~C.,  Yates J.~A.,  2001, MNRAS, 326, 1423

\bibitem[\protect\citeauthoryear{Ward-Thompson, Andr\'{e} \&
  Kirk}{Ward-Thompson et~al.}{2002}]{ward-thompson.et.al02}
Ward-Thompson D.,  Andr\'{e} P.,    Kirk J.~M.,  2002, MNRAS, 329, 257

\bibitem[\protect\citeauthoryear{Whitworth \& Bate}{Whitworth \&
  Bate}{2001}]{whitworth&bate02}
Whitworth A.~P.,  Bate M.~R.,  2001, MNRAS, 333, 679

\bibitem[\protect\citeauthoryear{Whitworth \& Ward-Thompson}{Whitworth \&
  Ward-Thompson}{2001}]{whitworth&wardthompson01}
Whitworth A.~P.,  Ward-Thompson D.,  2001, ApJ, 547, 317

\bibitem[\protect\citeauthoryear{Wilson \& Rood}{Wilson \&
  Rood}{1994}]{wilson&rood94}
Wilson T.~L.,  Rood R.~T.,  1994, ARA\&A, 32, 191

\end{thebibliography}
\end{document}